\documentclass[prd,twocolumn,superscriptaddress]{revtex4}
\usepackage{graphicx}                                                                                                                                                                                                                             
\usepackage{amsmath}
\usepackage{amsfonts}
\usepackage{amssymb}
\usepackage{mathrsfs}
\usepackage{textcomp}
\usepackage{graphics}
\usepackage{caption}
\usepackage[titletoc]{appendix}
 \usepackage{subfig}
 \usepackage[utf8]{inputenc}

\begin{document}

\begin{abstract}
We study quintessential inflation  with an inverse hyperbolic type potential
$V(\phi) = {V_0}/{\cosh \left( {\phi^n}/{\lambda^n} \right)}$,
where $V_0$, $\lambda$ and ``n'' are parameters of the theory. We obtain a  bound on
$\lambda$ for different values of the parameter n. The spectral index and the tensor-to-scalar-ratio
fall in the $1 \sigma$ bound given by the Planck 2015 data for $n \geq 5$ for certain values of $\lambda$.
However for $3 \leq n < 5$ there exist values of $\lambda$ for which the spectral index and the tensor-to-scalar-ratio
fall only within the $2 \sigma$ bound of the Planck data. Furthermore, we show that the scalar field with
the given potential can also give rise to late time acceleration if we invoke the coupling to massive neutrino matter. We also consider
the instant preheating mechanism with Yukawa interaction and put bounds on the coupling constants for our model using the
nucleosynthesis constraint on relic gravity waves produced during inflation.
\end{abstract}

\title{Quintessential Inflation in a thawing realization}
\author{Abhineet Agarwal}
\affiliation{Centre for Theoretical Physics, Jamia Millia Islamia,New Delhi-110025,India}
\author{R. Myrzakulov}
\affiliation{Eurasian International Center for Theoretical Physics, Eurasian National University, Astana 010008, Kazakhstan}
\author{M. Sami}
\affiliation{Centre for Theoretical Physics, Jamia Millia Islamia,New Delhi-110025,India}
\author{Naveen K. Singh}
\affiliation{Centre for Theoretical Physics, Jamia Millia Islamia,New Delhi-110025,India}
\date{\today}

\maketitle

\section{Introduction}
The standard model of the Universe needs to be modified at early and late times in order to 
address the problems therein. It is interesting to note that both the early and late time shortcomings
of the model can  be successfully  circumvented by introducing an
early phase of accelerated expansion called inflation together with
late time cosmic acceleration. The inflationary paradigm
resolves the horizon problem, the flatness problem and the monopole
problem and provides a mechanism for generation of primordial
perturbations \cite{GuthInflation,Starobinsky:1980te,Starobinsky:1982ee,Linde:1983gd,Linde:1981mu,Gasperini:1992pa,Lyth:1998xn}. Late time cosmic acceleration is
a recent discovery \cite{Perlmutter:1998np,Riess:1998dv} backed up by observations of type Ia supernovae,
CMB background and galaxy clustering. It requires the presence of an
exotic matter of energy momentum tensor with a large negative
pressure dubbed dark energy\cite{QIPeebles, Dodelson:1999am,Turner:1999kz,Amendola:2000uh,Caldwell:1999ew,Peebles:2002gy, QIRGWBSamiSahni,Copeland:2006wr}.
Cosmological constant and slowly rolling scalar fields provide a
viable example of dark energy; the effect can also be mimicked by
large scale modification of gravity. The presence of a late time phase
of accelerated expansion is essential in order to resolve the age crisis \cite{Krauss:1995yb} which arises in the hot big bang 
model. Clearly, accelerated expansion plays an important role in the history of our Universe$-$
the big bang model is sandwiched between two phases of accelerated
expansion

To unify  these two important concepts of inflation and late time acceleration, the notion
of ``Quintessential Inflation'' was
introduced where the inflaton field behaves like quintessence during late times \cite{QIPeebles,Giovannini:1999qj,Giovannini:1999bh,Giovannini:2003jw}. A major problem with
quintessential inflation is the ``cosmic coincidence'' problem which occurs due the fact that the energy density of the scalar field
and the matter energy density have comparable values today and for this to occur the conditions in the early universe have to be
very precisely predetermined. This tight constraint on the initial conditions can be relaxed if we introduce the notion of the tracker
field whose equation of motion has attractor type behavior\cite{TrackerSteinhardt, QIPlanck2015SamiWali,QIRGWBSamiSahni,RGWBISamiSahni}. 
On the contrary, the thawing dynamics has a strong dependence on the initial values. However,  as pointed out by Weinberg \cite{Weinberg:2000yb},
the scalar field irrespective of its behaviour can not address the fine tuning problem associated with the cosmological 
constant. Indeed,  replacing the cosmological constant by a scalar field translates the cosmological constant problem  into the problem of naturalness
of the scalar field.

At the onset, it sounds quite plausible to unify the early and late
time evolution using a single scalar field that does not disturb the
thermal history of Universe. For instance, a field, whose potential
is initially shallow at early times followed by a steep behaviour and then finally shallow again
at late times, should full fill the said requirement. Unfortunately,
the generic potentials do not change their shape that frequently$-$
they are either shallow at early times followed by a steep behaviour
thereafter or vice-versa. In the first category, one can realize
inflation  followed by a scaling solution; a suitable mechanism is then required to
exit from the scaling regime in order to obtain late time acceleration. In the second
category, one requires extra damping at early times to facilitate
inflation. Such an effect, in particular, could be induced by high
energy corrections on the Randall-Sundrum brane. Unfortunately, this
scenario is ruled out by observation$-$ the tensor to scalar ratio
of perturbations is too high in this case. However, the first
category of models could  give rise to a viable scenario of
quintessential inflation.

Let us note that in the first category of models, one can  exit from the scaling regime to
obtain late time acceleration by coupling the field non minimally to matter. The coupling  builds up
dynamically in the matter era giving rise to a minimum in the 
potential where the field could settle down leading to a de Sitter like solution. 
However, coupling to cold dark matter might destroy the matter phase as the minimum in this case would be induced as soon as the matter phase is established whereas there is no such issue associated
with the non-minimal
coupling of the field to massive neutrino matter. On the other hand, neutrinos become non-relativistic at  late times
and as a result the coupling of massive neutrinos to the scalar fields becomes more prominent. The latter ensures the 
appearance of a minimum in the field potential at late times \cite{QIPlanck2015SamiWali,Hossain:2014xha, Amendola:2007yx} which 
is desirable for the commencement of late time cosmic acceleration. In this paper we follow in the footsteps of 
the above authors and discuss a model with inverse cosine-hyperbolic potential that might successfully 
give rise to quintessential inflation.

The inverse cosine-hyperbolic potential is a  tachyonic type potential  and is inspired from 
Sring theory \cite{Lambert:2003zr,Djordjevic:2016pdh}.   Some of the works related to tachyonic potential can be found in 
Refs. \cite{Sen:1999mg,Sen:2002nu}. The
inverse cosine hyperbolic type of potential is mentioned in the Refs.  \cite{Lambert:2003zr,Djordjevic:2016pdh}. However, it is not studied 
in detail by considering the cosmological data of Planck 2015. Further, the late time evolution of the Universe
is  not studied yet. In this paper we  generalize the inverse cosine hyperbolic potential to describe 
quintessential inflation.

\section{Inflation and observational constraints}
As mentioned in the introduction, we are looking for a scenario that would facilitate slow roll at early times followed by scaling behavior in the post inflationary regime such that the exit to late time acceleration is caused by non-minimally coupled massive neutrinos. To this effect, we shall 
consider the following action,
\begin{align}
\label{action}
S &= \int{d^4 x \sqrt{-g} \left[ \frac{M_{pl}^2}{2} R - \frac{1}{2} {\partial^{\mu} \phi} {\partial_{\mu} \phi} - V (\phi) \right]}
+S_m  \nonumber \\ & +S_{\nu}\left(\mathcal{C}^2(\phi) g_{\mu\nu};\psi_{\nu}\right)+ S_R.
\end{align}
with,
\begin{eqnarray}
\label{ourpot}
&& V(\phi) = \frac{V_0}{\cosh \left( \frac{\phi^n}{\lambda^n} \right)}= \frac{V_0}{\cosh[{\beta^n (\phi/M_{pl})^n}]}\\
&& \mathcal{C}\sim e^{\gamma \phi/M_{pl}}
\end{eqnarray}
where $\lambda = \alpha M_{pl}$; $\beta = \frac{1}{\alpha}$ and $\alpha$, $\beta$ are dimensionless parameters and $n$ is a positive integer.  

Eq. (\ref{action}) includes the actions for standard matter ($S_m$), massive neutrino ($S_\nu$), and radiation ($S_R$). We have assumed that the scalar field is directly coupled to massive neutrino matter whereas the dark matter is minimally coupled. As for inflation, a remark about the matter actions is in order. Since matter is generated during reheating, the said actions should be dropped while disusing inflation.
 
In what follows, we specialize
to  a flat FRW background  to obtain the evolution equations for the  action (\ref{action}),
\begin{eqnarray}
&& 3 H^2 M_{pl}^2 =
\frac{1}{2} {\dot{\phi}}^2 + V (\phi) \label{Frone},\\
&& \left( 2 \dot{H} + 3 H^2 \right) M_{pl}^2 = - \frac{1}{2} {\dot{\phi}}^2 + V (\phi) \label{Frtwo}
\end{eqnarray}
and
\begin{equation}
\ddot{\phi} + 3 H \dot{\phi} + \frac{d V}{d \phi} = 0.
\end{equation}

The slow roll parameters for a potential $V(\phi)$ are defined as usual \cite{QIPlanck2015SamiWali,LythontheHilltop},
\begin{equation}
\epsilon = \frac{M_{pl}^2}{2} {\left( \frac{1}{V} \frac{d V}{d \phi} \right)}^2 \label{epsilon},
\end{equation}
and
\begin{equation}
\eta = \frac{M_{pl}^2}{V} \frac{d^2 V}{d {\phi}^2} \label{eta}.
\end{equation}
The end of inflation is marked by,
\begin{equation}
\epsilon |_{\phi = \phi_{end}} = 1 \label{end} ,
\end{equation}
where ``end'' represents the value at the end of inflation.
Let us consider a period which begins when the modes cross the horizon and ends with the end of inflation. Then the number of e-foldings
during this period is given by \cite{QIPlanck2015SamiWali,LythontheHilltop},
\begin{align}
N &= M_{pl}^{-1} \int_{\phi_{end}}^{\phi}{\frac{d \phi}{\sqrt{2 \epsilon(\phi)}}} \\
&= \frac{1}{M_{pl}^2} \int_{\phi_{end}}^{\phi}{\frac{V(\phi')}{V'(\phi')}d \phi'} \label{efoldings}  .
\end{align}
The tensor to scalar ratio $r$ is given by,
\begin{equation}
r = 16 \epsilon \label{ttsr},
\end{equation}
and the scalar spectral index $n_s$, which is defined as,
\begin{equation}
n_s - 1 = \frac{d(log P_R)}{d(log k)},
\end{equation}
where $P_R$ is the spectrum of curvature perturbations, is reduced to the form,
\begin{equation}
n_s = 2 \eta - 6 \epsilon + 1 \label{spectralindex}.
\end{equation}
We now study a model based on a potential given by Eq. (\ref{ourpot}) for general $n$
%
and  derive expressions for the slow roll parameters and the spectral index in terms of $n$. Introducing a
dimensionless scalar field $\chi = \frac{\phi}{M_{pl}}$ and
using Eq.~(\ref{epsilon}), Eq.~(\ref{eta}) and Eq.~(\ref{ourpot}), we obtain,
\begin{equation}
\epsilon = \frac{1}{2} n^2 \beta^{2n} \chi^{2n - 2} \tanh^2 \left( \beta^n \chi^n \right) \\ \label{epsilonthree},
\end{equation}
\begin{equation}
r = 16 \epsilon = 8 n^2 \beta^{2n} \chi^{2n - 2} \tanh^2 \left( \beta^n \chi^n \right) \\ \label{ttsrthree},
\end{equation}
and

\begin{widetext}
\begin{equation}
\eta = - n \beta^2 {\left(\beta \chi \right)}^{n - 2}
\Big[
n {\left( \beta \chi \right)}^n
+ (n - 1) \tanh \left( \beta^n \chi^n \right)
- 2n {\left( \beta \chi \right)}^{n} \tanh^2 \left( \beta^n \chi^n \right)
\Big] \label{etatwo}  .
\end{equation}
\end{widetext}

\newpage

From Eq.~(\ref{spectralindex}), Eq.~(\ref{epsilonthree}) and Eq.~(\ref{etatwo}) we also obtain the spectral index

\begin{widetext}
\begin{equation}
n_s = 1 - 2 n^2 \beta^{2n} \chi^{2n - 2} - 2n (n - 1) \beta^n \chi^{n - 2} \tanh \left( \beta^n \chi^n \right)
+ n^2 \beta^{2n} \chi^{2n - 2} \tanh^2 \left( \beta^n \chi^n \right) . \label{ns}
\end{equation}
\end{widetext}

Eq.~(\ref{end}) gives us
\begin{equation}
\frac{n}{\sqrt{2}} \beta^n \chi^{n - 1} \tanh \left( \beta^n \chi^n \right)|_{\chi= \chi_{end}} = 1. \label{constraint1}
\end{equation}

From Eq.~(\ref{efoldings}) we obtain,

\begin{equation}
N_k = \frac{1}{n \beta^n}
\int_{\chi}^{\chi_{end}}{\frac{\coth \left( \beta^n \chi^n \right)}{\chi^{n - 1}} d \chi}.
\end{equation}

We now set the number of e-foldings to be $60$. Comparing the theoretical value of the power spectrum,
\begin{equation}
P_{R} = \frac{1}{24 \pi^2 M_{pl}^4} \frac{V}{\epsilon},
\end{equation}
  \begin{figure}[h]
 \centering
 \includegraphics[scale=.7]{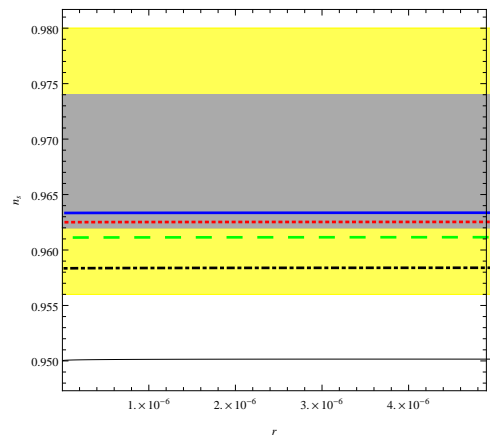}
 \caption{The above parametric plot shows the variation of the spectral index $n_s (\beta)$ with respect to the  tensor-to-scalar ratio
 $r (\beta)$ for different values of n with $\beta$ lying in the range 0.2 - 7.19  where n takes values from 2 to 6.
 For $n$ is equal to $2,3,4,5,6$; we obtain $\chi_{end}$ is equal to $17.68,7.68,6.09,5.55,5.29$ and $\chi_{i}$ is equal to
 $1.80,2.18,2.55,2.83,3.04$ respectively for $\beta=0.2$. For $n$ is equal to $2,3,4,5,6$; we obtain $\chi_{end}$ is equal to $0.06,0.08,0.09,0.097,0.1$ and $\chi_{i}$ is equal to
 $0.001,0.01,0.02,0.03,0.04$ respectively for $\beta=7.19$. Both $\chi_{i}$ and $\chi_{end}$ decrease monotonically with increase 
 in $\beta$ for all $n$.
 The plot also shows
 the $1-\sigma$ (grey shaded area) and $2-\sigma$ ( yellow shaded area) intervals as given by Planck 2015. The plot shows that
 $n = 3$ (black dotted-dashed) and $n = 4$ (green-dashed) lie in the $2 - \sigma$ interval, $n = 5$ (red-dotted) and $n = 6$ (blue-solid) lie in the
 $1 - \sigma$ interval and $n = 2$ (black-thin) does not lie even in the $2 -\sigma$ interval.}
 \label{allplots}
 \end{figure}
where $M_{pl} = {\left( 8 \pi G_{N} \right)}^{- \frac{1}{2}} = 2.4 * 10^{18} \mbox{GeV}$, with its observational value
$P_R \approx 2.19 * 10^{-9}$ and combining it with Eq.~(\ref{ourpot}) and Eq.~(\ref{epsilonthree}), one derives an estimate for  $V_0$, 

\begin{equation}
V_0 = \left( 2.14916 *10^{68} \right) \beta^{2n} \chi_i^{2n - 2}
\cosh \left( \beta^n \chi_i^n \right) \tanh^2 \left( \beta^n \chi_i^n \right).
 \end{equation}
 where, $\chi_i$ is the value of $\chi$ at the beginning of inflation.
%
 \begin{figure}[]
\begin{minipage}{.7\linewidth}
\centering
\subfloat[]{\label{variation1}\includegraphics[scale=.6]{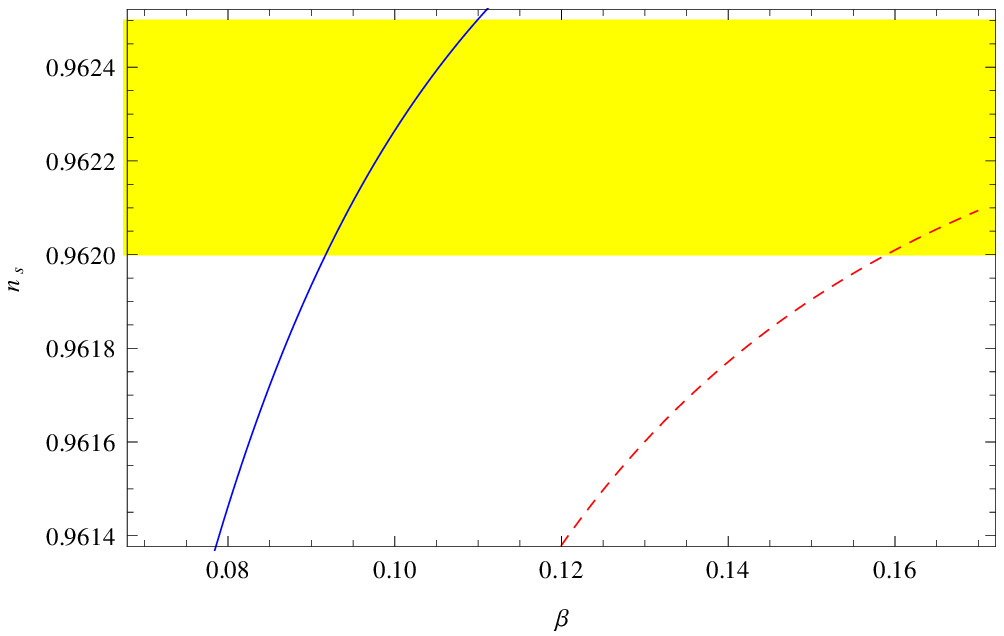}}
\end{minipage}\par\medskip
\begin{minipage}{.7\linewidth}
\centering
\subfloat[]{\label{variation2}\includegraphics[scale=.6]{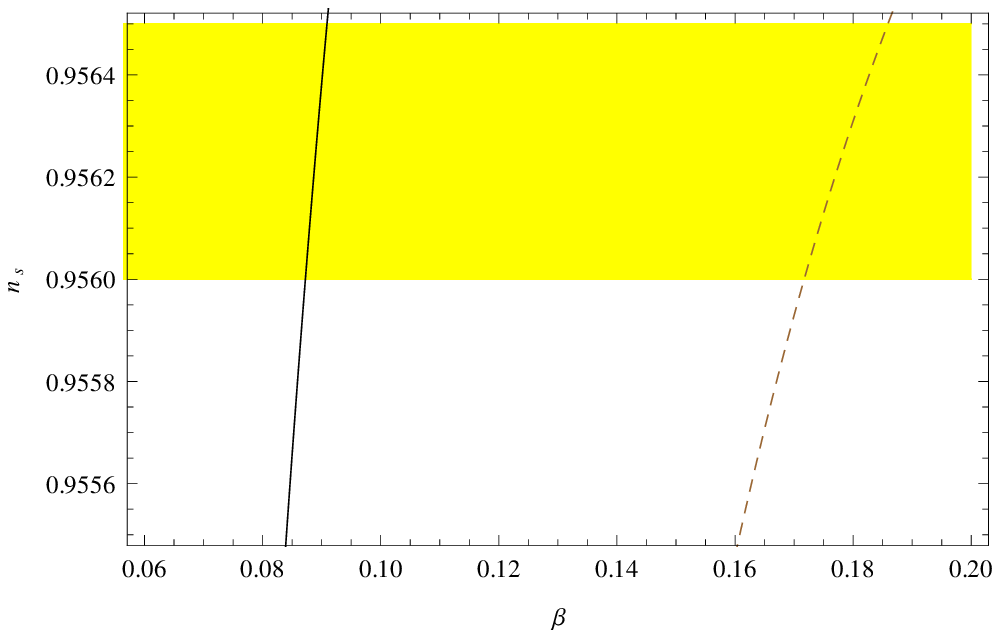}}
\end{minipage}

\caption{The above plot shows how $n_s$ varies with respect to $\beta$ for $n = 6$ (blue), $n = 5$ (red-dashed), $n = 4$ (black)
 and $n = 3$ (brown-dashed) for low values of $\beta$. In sub-figure (\ref{variation1}), the yellow
 coloured shade indicates the $1-\sigma$ region with a lower boundary given by $n_s =0.962 $ and in sub-figure
 (\ref{variation2}), it indicates the $2-\sigma$ region with a lower boundary given by $n_s =0.956 $.}
  \label{lowbeta}
\end{figure}

According to the Planck 2015 results \cite{planck2015}, $n_s = 0.968 \pm 0.006$ at the $1 \sigma$ confidence level.  For $n=1$,
Eqns.~(\ref{constraint1}) and (\ref{ns}) reduce to
\begin{equation}
 \epsilon = \frac{\beta^2}{2} {\left( \tanh \left( \beta \chi_{end} \right) \right)}^2 = 1
\end{equation}
and
\begin{equation}
 n_s= 1 - \beta^2 \left[ 2 - \frac{2 e^{- 2 N \beta^2}}{\beta^2 - 2 + 2 e^{-2 N \beta^2}} \right],
\end{equation}
respectively, where the  expression for $n_s$ is derived analytically by inverting the equation,
\begin{equation}
 N(k) = \frac{1}{\beta^2}
 \left[ \ln \left( \sqrt{\frac{2}{\beta^2 - 2}} \right)
- \ln \left| \sinh \left( \beta \chi \right) \right| \right]
\end{equation}
to obtain $\sinh{(\beta \chi})$ and hence $\tanh{(\beta \chi})$. Since $\tanh{(\beta \chi_{end})}$ can take a maximum value of one,
therefore, for inflation to end and  hence for $\epsilon$ to be unity, $\beta \geq \sqrt{2}$. However, for $\beta = \sqrt{2}$,
$n_s=-1$ and as the value of $\beta$ increases further $n_s$ becomes more and more negative.
 For $n=2$, we have a positive value of $n_s$ but it still does not fall in the observational range as shown
in Fig. (\ref{allplots}).  As we increase the value of $n$, the theoretical value of $n_s$ shifts towards the observed value.
From  Fig. (\ref{allplots}) it is clear that for $n = 3$ and  $4$, there exist no values of $\beta$ for which $n_s$ lies in this
range at the $1 \sigma$ confidence level. However for $\beta \geq 0.17$, $n = 3$ and $\beta \geq 0.085$, $n = 4$ (Fig. (\ref{lowbeta})) the value
of $n_s$ given by the theoretical model agrees with the Planck 2015 results up to the
$2 \sigma$ confidence level. Interesting results are found for $n=5$ and $6$ for which the plots
are shown in   Fig. (\ref{allplots}). For $n=5$, $\beta \geq 0.16$  and
$n=6$, $\beta \geq 0.09$ (Fig. (\ref{lowbeta})) the theoretical value of the spectral index $n_s$ is in  agreement with the Planck 2015 data
 up-to the $1 \sigma$ confidence level. For $\beta < 0.9$, we have $\phi_f > \phi_i > M_{pl}$ whereas
$\beta \gtrsim 0.9$,  $\phi_i < \phi_f < M_{pl}$ for $n=5$ and $6$. For higher value of $\beta$, $\phi$ becomes much smaller than
 $M_{pl}$ and also the spectral index $n_s$ attains a constant value of $0.9625$
and $0.9633$ respectively for $n=5$ and $6$. The tensor-to-scalar ratio $r$  satisfies the Planck
2015 data \cite{planck2015}, i.e., $r<0.1$  for all the values of $n$ discussed here. \\

We should emphasize that the permitted range of parameters  in the model leads to small 
numerical values of $r$, thereby, the Lyth bound \cite{ala} is easily satisfied in the sub-Planckian region.
Indeed, using Eq. (\ref{efoldings}), 
\begin{equation}
 N=\frac{1}{M_{pl}} \int^{\phi_{in}}_{\phi_{end}} \frac{d \phi}{\sqrt{2 \epsilon}}
 \to N\lesssim \frac{|\phi_{in}-\phi_{end}|}{M_{pl}\sqrt{2 \epsilon_{min}}}\equiv \frac{\delta \phi}{M_{pl}\sqrt{2 \epsilon_{min}}},
\end{equation}
As $\epsilon$ is a monotonically increasing function in the model under consideration,  $\epsilon_{min}= \epsilon_{in}=r/16$ and the range of inflation is given by,
$\delta \phi \gtrsim N \sqrt{2 \epsilon_{in}} M_{pl}=N\sqrt{r/8} M_{pl}$. For $r\sim 10^{-6}$, $N=60$, we find that $\delta\phi\gtrsim .02 M_{pl}$ and Lyth bound is satisfied as stated.

\begin{figure*}
\begin{minipage}{.47\linewidth}
\centering
\subfloat[]{\label{main:a}\includegraphics[scale=.8]{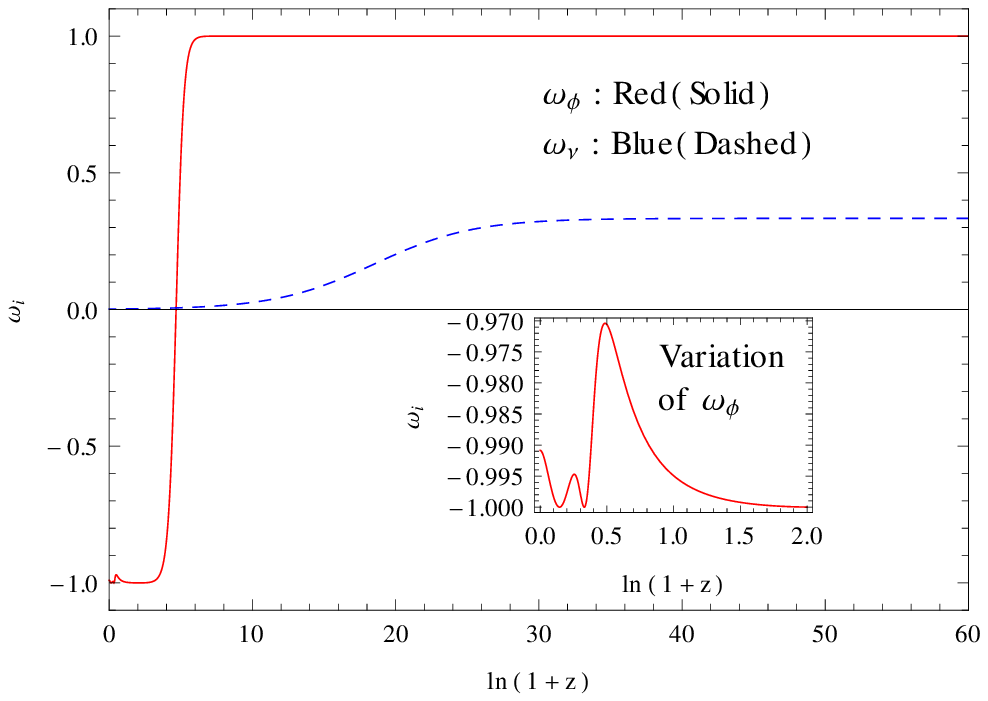}}
\end{minipage}%
\begin{minipage}{.47\linewidth}
\centering
\subfloat[]{\label{main:b}\includegraphics[scale=.8]{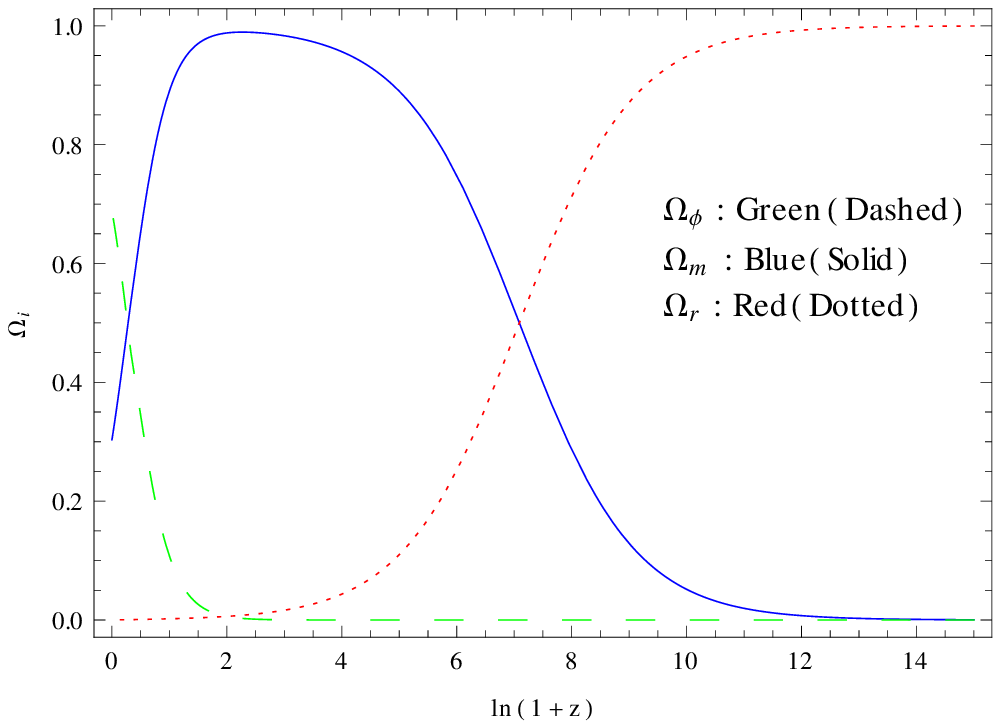}}
\end{minipage}\par\medskip
\begin{minipage}{.47\linewidth}
\centering
\subfloat[]{\label{main:c}\includegraphics[scale=.8]{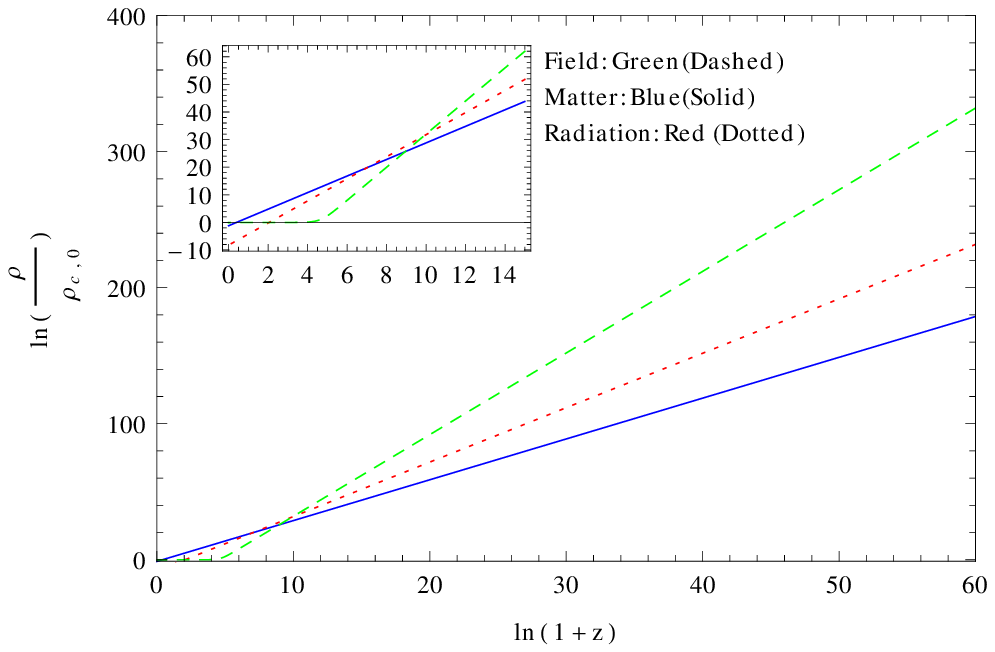}}
\end{minipage}%
\begin{minipage}{.47\linewidth}
\centering
\subfloat[]{\label{main:d}\includegraphics[scale=.8]{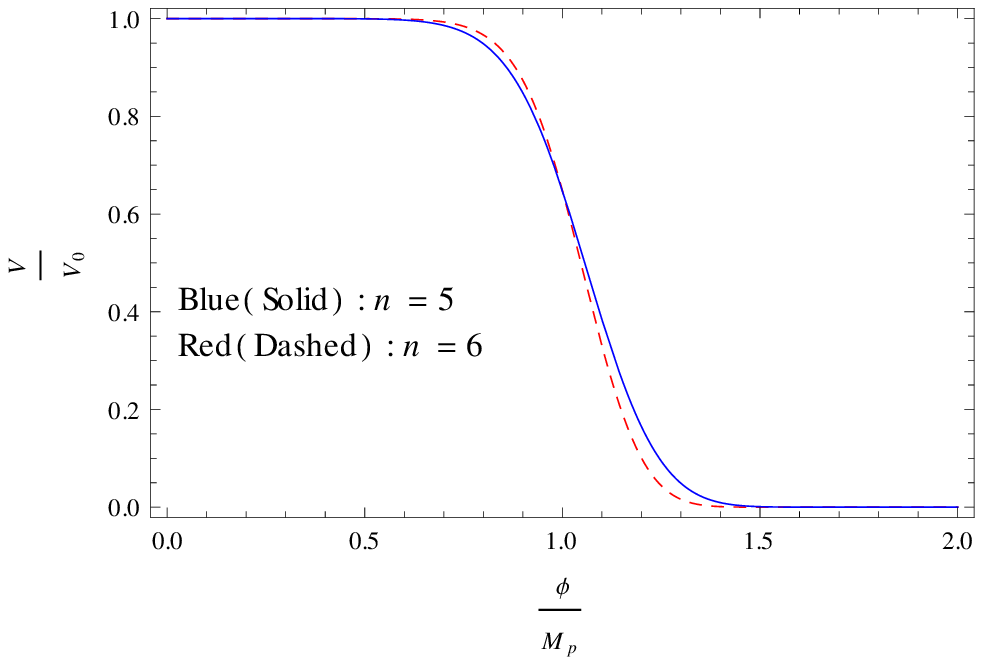}}
\end{minipage}\par\medskip
\caption{ $\omega_{i}$ ,  $\Omega_{i}$ and  $\ln(\rho/\rho_{c,0})$
are plotted as a function of $\ln{(1+z)}$ in the sub-figures (\ref{main:a}), (\ref{main:b}) and (\ref{main:c})
  respectively  for $n=6$.  The normalized potential $\frac{V}{V_0} = \frac{1}{\cosh \big[ \beta^n \left(\phi/M_{pl}\right)^n\big]}$ is
  plotted for $n=5$ and $n=6$ using the blue and red-dashed lines respectively in the sub-figure (\ref{main:d}).}
  \label{evolution}
\end{figure*}
\section{Preheating based upon instant particle production}
In the paradigm of quintessential inflation, the underlying potential is typically of a run away type. Thereby the standard mechanism of (p)reheating\cite{LDH,BPS,Starobinsky,Starobinsky2} can not be implemented in this case. One
of the possible alternatives is provided by gravitational particle production. However, this process is inefficient and as a result the field spends a 
long time in the kinetic regime ($\rho_\phi\sim 1/a^6$). The energy density of gravity waves which were produced at the end of
inflation becomes sufficiently larger than  $\rho_\phi$ during the kinetic regime and this leads to a violation of the nucleosynthesis constraint. The nucleosynthesis 
constraint restricts the duration of the kinetic regime and we can also put a lower bound on the  reheating
temperature using this constraint. Therefore, we require an efficient reheating mechanism to address the issue. The instant 
preheating mechanism turns out to be suitable. 

 In what follows, we shall consider a scenario where
the inflaton field $\phi$ interacts with another scalar field $\chi$ and we assume that the scalar
field $\chi$ interacts with the matter field  $\psi$ such that\cite{Campos1},
\begin{equation}
\phi \quad \xrightarrow{g} \quad \chi \quad \xrightarrow{h} \quad \psi \bar{\psi}.
\end{equation}
 Let us choose the following interaction terms in the  Lagrangian,
\begin{equation}
\label{Lint}
\mathscr{L}_{int} = - \frac{1}{2} g^2 \phi^2 \chi^2 - h \bar{\psi} \psi \chi
\end{equation}
in order to implement the aforesaid. Here $g$ and $h$ are positive coupling constants \cite{SamiDadich,InstantPreheating,FermionPreheating,VariableGravity,ala}. The form of the 
Lagrangian is chosen such that $\chi$ field has no bare mass and its effective mass is given by
$m_{\chi} = g \left| \phi \right|$ \cite{InstantPreheating2}. Particle production commences after 
inflation provided that $m_\chi$ changes non-adiabatically, $\dot{m}_\chi\gtrsim m^2_\chi$. The mass
of produced $\chi$ particles grows as the field evolves to larger values after the end of inflation
and $\chi$ decays into various species of particles. Assuming that all the energy is instantaneously 
converted into radiation and thermalized, one can obtain the following estimate(see Appendix A for details),
\begin{equation}
\label{gconst}
{\left( \frac{\rho_{\phi}}{\rho_r} \right)}_{end} \simeq \frac{3}{2} \frac{(2 \pi)^3}{g^2}.
\end{equation}
As mentioned before, the energy density of gravity waves  becomes significantly larger than $\rho_\phi$ during the kinetic 
regime and this might later lead to a violation of the nucleosynthesis constraint during the radiative regime. The said constraint,
\begin{equation}
 \left(\frac{\rho_g}{\rho_r}\right)_{eq}\lesssim 10^{-2},
\end{equation}
where, $\left(\frac{\rho_g}{\rho_r}\right)_{eq}
 = \frac{64}{3 \pi} h_{GW}^2 \left(\frac{\rho_{\phi}}{\rho_r}\right)_{end}$ combined with the value of the dimensionless
 gravity wave amplitude $h_{G.W.}$ calculated for our model (see appendix A) 
puts an upper bound on $\rho_\phi/\rho_{r}$,

\begin{equation}
 \left(\frac{\rho_{\phi}}{\rho_r}\right)_{end} \lesssim 2.53 \times 10^{11} \label{CONSTRAINT}
\end{equation}
at the end of inflation. This gives us a lower bound on $(\rho_r)_{end}$, $(\rho_r)_{end} \gtrsim 2.48 \times 10^{49} GeV^4$ 
and hence a bound on the reheating temperature, $(T_r)_{end} \gtrsim 2.2 \times 10^{12} GeV$, is obtained. Using
 Eq. (\ref{gconst}) together with Eq. (\ref{CONSTRAINT}), we find out the lower limit on $g$. The limit on $h$ can be 
 set by avoiding the back reaction of $\chi$ particles on the post inflationary 
dynamics. We find a comfortable range in the parameter space $(g,h)$ which can give rise to  efficient preheating 
after inflation (see Appendix A for details).
\section{Late time acceleration with non-minimal coupling to massive neutrino matter}

In the absence of non-minimal coupling, the scalar field with a potential
given by Eq.~(\ref{ourpot}) can not give rise to late time acceleration  as the potential is steep. To obtain late time 
acceleration one needs to introduce a feature in the potential which can be achieved by  coupling
the scalar field non minimally to massive neutrino matter. The coupling of the scalar  field with neutrinos is not prominent in the radiation era while it  plays
a crucial role during the late stages of evolution where it induces a minimum in the potential giving rise 
to a de Sitter solution. The result is a late time attractor. As mentioned before, it is desirable to leave the matter era 
intact and assume that the field couples to massive neutrino matter only.

The neutrinos here are relativistic $\left( p_{\nu} = \frac{\rho_{\nu}}{3} \right)$
like radiation during early times but turn non relativistic $\left( p_{\nu} = 0 \right)$ at late times. 
This behaviour can be mimicked  by   the following ansatz for parametrization of
$\omega_{\nu} (z)$,
\begin{equation}
\omega_{\nu} (z) = \frac{p_{\nu}}{\rho_{\nu}}
= \frac{1}{6} \left( 1 + \tanh \left[ \frac{\ln (1 + z) - z_{eq}}{z_{dur}} \right] \right) .
\end{equation}
The equation of state $\omega_{\nu} (z)$ is thus chosen keeping in mind the
phase transition from the relativistic state to the non relativistic one.
$z_{eq}$ and $z_{dur}$ give the time and duration of the transition.
The equation of continuity for massive neutrinos, in presence of coupling, takes the form,
\cite{QIPlanck2015SamiWali,ala}.
\begin{equation}
\dot{\rho}_{\nu} + 3 H \left( \rho_{\nu} + p_{\nu} \right)
= \gamma \left( \rho_{\nu} - 3 p_{\nu} \right) \frac{\dot{\phi}}{M_{pl}} . \label{rhonueqn}
\end{equation}
 Varying the action (\ref{action}) with respect to $\phi$ and following the Ref. \cite{ala}, we obtain,
\begin{equation}
\ddot{\phi} + 3 H \dot{\phi} = - \frac{\partial V}{\partial \phi}
- \gamma \left( \rho_{\nu} - 3 p_{\nu} \right) . \label{phieqn}
\end{equation}
 We note that in the radiation era when the neutrino is relativistic, in Eqs (\ref{rhonueqn}) and (\ref{phieqn}),
the coupling term of the scalar field and neutrinos vanishes since
$\rho_{\nu}-3 p_{\nu}=0$. Thus the coupling becomes effective only when the neutrinos become non-relativistic. We thus have a system of equations,
\begin{equation}
3 H^2 M_{pl}^2 = \frac{1}{2} \dot{\phi}^2 + V(\phi) + \rho_m + \rho_r + \rho_{\nu},
\end{equation}
\begin{equation}
(2 \dot{H} + 3 H^2) M_{pl}^2 = - \frac{1}{2} \dot{\phi}^2 + V(\phi) - \frac{1}{3} \rho_r - p_{\nu},
\end{equation}
\begin{equation}
\ddot{\phi} + 3 H \dot{\phi} = - \frac{\partial V}{\partial \phi}
- \gamma \left( \rho_{\nu} - 3 p_{\nu} \right),
\end{equation}
\begin{equation}
\dot{\rho}_{\nu} + 3 H \left( \rho_{\nu} + p_{\nu} \right)
= \gamma \left( \rho_{\nu} - 3 p_{\nu} \right) \frac{\dot{\phi}}{M_{pl}} .
\end{equation}

\begin{equation}
\dot{\rho}_i + 3 H \left( \rho_i + p_i \right) = 0 ; \ \ i = \mbox{matter and  rad.},
\end{equation}
which we can solve to obtain the evolution of the various energy densities. To solve this system of differential equations,
we introduce the dimensionless variables $x$, $y$, $\omega_{\nu}$, $\Omega_m$, $\Omega_r$ and $\lambda$ where
$x = \frac{\dot{\phi}}{\sqrt{6} H M_{pl}}$, $y = \frac{\sqrt{V}}{\sqrt{3} H M_{pl}}$,
$\Omega_m = \frac{\rho_m}{3 H^2 M_{pl}^2}$, $\Omega_r = \frac{\rho_r}{3 H^2 M_{pl}^2}$,
  $\Omega_{\phi} = \frac{\rho_{\phi}}{3 H^2 M_{pl}^2}$ and $\lambda = - M_{pl} \frac{V,_{\phi}}{V}$.
The evolution  equations can then be cast in a convenient form,
\begin{eqnarray}
x' &=& \frac{x}{2} \Big[ 3 \left( \omega_{\nu} - 1 \right) + \left( 1 - 3 \omega_{\nu} \right) \Omega_r
- 3 \omega_{\nu} \Omega_m \nonumber \\
&-& 3 \omega_{\nu} x^2 - \left( 3 \omega_{\nu} + 3 \right) y^2 \Big]
+ \frac{3}{2} x^3 + \sqrt{\frac{3}{2}} \lambda y^2 \nonumber \\
&+& \sqrt{\frac{3}{2}} \left( 3 \omega_{\nu} - 1 \right) \gamma 
\left( 1 - \Omega_m - \Omega_r - x^2 - y^2 \right) , \nonumber \\ \\
y' &=& \frac{y}{2} \Big[ 3 \left(  \omega_{\nu} + 1 \right) + \left( 1 - 3 \omega_{\nu} \right) \Omega_r
- 3 \omega_{\nu} \Omega_m \nonumber \\
&+& 3 \left( 1 -  \omega_{\nu} \right) x^2 - 3 \omega_{\nu} y^2 - \sqrt{6} x \lambda \Big] - \frac{3}{2} y^3 , \\
\Omega_r ' &=& - \Omega_r \Big[ \left( 1 - 3 \omega_{\nu} \right) + \left( 3 \omega_{\nu} - 1 \right) \Omega_r
+ 3 \omega_{\nu} \Omega_m \nonumber \\
&+& 3 \left( \omega_{\nu} - 1 \right) x^2 + 3 \left( \omega_{\nu} + 1 \right) y^2 \Big],
\end{eqnarray}
\begin{eqnarray}
\Omega_m' &=& \Omega_m \Big[ 3 \omega_{\nu} + \left( 1 - 3 \omega_{\nu} \right) \Omega_r
- 3 \omega_{\nu} \Omega_m \nonumber \\
&+& 3 \left( 1 - \omega_{\nu} \right) x^2 - 3 \left( \omega_{\nu} + 1 \right) y^2 \Big], \\
 \omega_{\nu}' &=& \frac{2 \omega_{\nu}}{z_{dur}} \left(3 \omega_{\nu} - 1\right) ,\\
z' &=& \sqrt{6} x .
\end{eqnarray}
Here, prime $'$ is derivative w.r. to $N$ and we have used the constraint, $\Omega_{\nu} = 1 - \Omega_r - \Omega_m - x^2 - y^2$.
 We now solve this dynamical system numerically for $\beta = 1$ and plot
these variables taking the initial conditions at the end of inflation. The initial values of $x$, $y$, $z$, $\omega_{\nu}$ $\Omega_m$ and
$\Omega_r$ are approximately $10^{-51}$, $10^{-51}$, $0.88$, $0.3$, $10^{-23}$ and $1$ respectively. $z_{dur}$ 
and $\gamma$ are $6.9$ and $100$ respectively. The initial value of $\Omega_{\nu}$ is $10^{-7}$. From the plots of
Fig. (\ref{evolution}), it is clear that we can obtain late time acceleration with neutrino coupling  using
these initial conditions, since we  can obtain the current values of $\Omega_{\phi}$, $\Omega_m$ and $\omega_{\phi}$ which
are approximately $0.7$, $0.3$ and $-1$ respectively. Similar graphs were plotted for   $\beta=2,3$ etc. and the same result was obtained. \\
A comment regarding the thawing nature of evolution is in order. The potential (\ref{ourpot}) belongs to the 
class of tracker potentials. 

Indeed, One can easily check that our potential behaves as $V \sim Exp(-\beta^n \phi^n/M^n_p)$ \cite{Geng:2017mic} asymptotically 
for large values of  $\phi$. However, on the contrary, we see a thawing
behaviour. It is clear from subfigure (\ref{main:d}) that after inflation ends, the potential drops sharply
resulting in a deep overshoot of $\rho_{\phi}$ with respect to the background energy density. 
Clearly, it takes long time for the field to recover from 
freezing which happens when the background energy density become comparable to $\rho_{\phi}$. When it happens, the 
field has already evolved to the slow roll regime near the minimum of the potential and consequently,
  tracker regime is missed out. 
  
  In this case nucleosynthesis poses no constraint on the slope of the potential unlike the case of standard exponential potential.
\begin{figure}
  \centering
 \includegraphics[scale=.8]{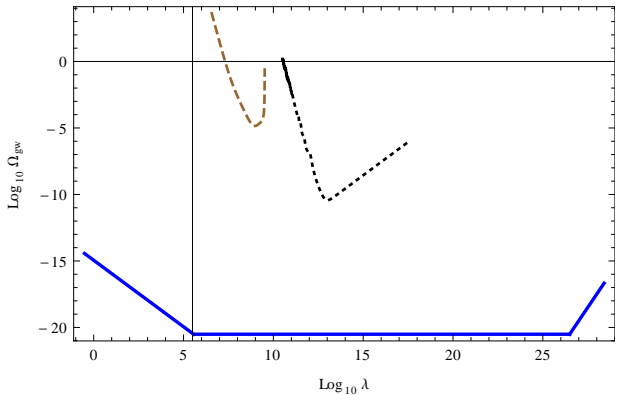} 
  \caption{The energy spectrum of the relic gravity waves is plotted as function of wavelength in the kinetic, radiation 
  and matter regimes together with the sensitivity curves of LIGO (brown-dashed) and LISA (black-dotted). The picture corresponds to the numerical value of reheating temperature, $T_{r(end)}=2.2\times 10^{12}GeV$; $\beta=1, N=60$ and $n=6$.}
  \label{lisaligo}
 \end{figure}

 \section{Conclusions}
 In this paper, we have considered a single scalar field model  with a potential $V_0/\cosh{\left(\phi^n/\lambda^n\right)}$ and demonstrated that the model can give
 rise to quintessential inflation. First, we have demonstrated the viability of the theory for inflation. In particular, we have shown that for lower  values of  $n$, the model is ruled out by observation. For instance if $n=2$, the theoretical value of $n_s$  fails to lie in the
 observed range. However, situation improves for higher values of $n$; potential  flattens significantly in this case. In particular, for $n=3$ and $4$, the considered potential provides
 a value of $n_s$ such that it can lie in the range of $2- \sigma$ confidence level of Planck 2015 data. The value of $\beta$ is greater than $0.17$ and
 $0.085$  for $n=3$ and $4$ respectively. The most profound results are obtained for $n=5$ and $6$. The theoretical value of $n_s$ satisfies
 the  Planck 2015 data at the $1 \sigma$ 
confidence level and its value becomes $\sim 0.9625$ and  $\sim 0.963$ for $\beta  \geq0 .16$ and
 $\beta \geq 0 .09$ for  $n=5$ and $6$ respectively. It should be noted that our model mimics the hill-top potential \cite{LythontheHilltop} for
 $\beta \sim 1$ as $\phi<<M_p$. It is therefore not surprising that  the tensor-to-scalar ratio $r$  is negligibly small 
 in our case similar to that of the Starobinsky-model \cite{Starobinsky:1980}.  Nevertheless, the very small value of the tensor-to-scalar
 ratio is a salient feature of our model compared to other models discussed in the literature.
 If we take $\beta$ to be much less than unity, we find that the value of $r$ is significantly greater than $10^{-6}$. Further, we
 have  plotted the evolution of normalized values  of
 different components of the energy densities $\Omega_i$, $\omega_{\nu}$, $\omega_{\phi}$ and the energy density of the scalar field $\rho_{\phi}$.
 The plots explain that the universe undergoes late time acceleration, i.e., for today, $\Omega_{\phi} \sim0.7$, $\Omega_{m} \sim 0.3$ and
 $\omega_{\phi} = -1$. Though the underlying potential belongs to the tracker class, the thawing behaviour manifests itself in the post-inflationary
 region. This behaviour depends on the specific initial conditions taken at the end of inflation. The latter
 then leads to a deep overshoot of the field and  consequently misses the tracker region of the potential.
 Thus, in the model under consideration (in a thawing realization), the scalar field which drives inflation in
 the early Universe is also responsible for late time acceleration once we couple the scalar field to massive neutrino matter. Finally, we considered the instant preheating mechanism for
  particle production where  we assumed  that the scalar field couples to the matter field via Yukawa interaction. We have
 put bounds on the coupling constants $g$ and $h$ defined in the Yukawa interaction which are found
 to be $g \geq 3.8 \times 10^{-5}$ and $h \geq 2.5 \times  10^{-3}/\sqrt{g}$ such that $(T_r)_{end} \geq 2.2 \times 10^{12} GeV$.
 These values satisfy the nucleosynthesis constraint.  We have also studied the spectrum of relic gravity waves in our model (see appendix B); the generic feature of 
 the scenario includes the blue spectrum of gravity waves on scales smaller than comoving horizon scales. Fig. (\ref{lisaligo})
 shows the energy spectrum for   $n = 6$ and $\beta = 1$ as a function of the wavelength $\lambda$ together with the sensitivity 
 curves of advLIGO and LISA.
 \vspace{-.7cm} 
 \section*{Acknowledgement}
 Naveen K. Singh is thankful to the D.S. Kothari postdoctoral fellowship of University Grant Commission, India for the financial
 support. His fellowship number is F.4-2/2006 (BSR)/PH/14-15/0034. Abhineet Agarwal wishes to thank Md. Wali Hossain and Safia Ahmad for
 useful discussions.
 \vspace{-.3cm} 
 \section*{\large Appendices}
 \begin{appendices}
 \chapter{\hspace{1cm}\bf Appendix A: Instant preheating} \label{appxA}
\vspace{.5cm} \\
In order to implement the instant particle production, we consider the following Lagrangian,

\begin{equation}
\mathscr{L}_{int} = - \frac{1}{2} g^2 \phi^2 \chi^2 - h \bar{\psi} \psi \chi, \label{Lint1}
\end{equation}
where $g$ and $h$ are positive coupling constants with a restriction, $g,h<1$ such that a perturbative treatment is viable for the Lagrangian (\ref{Lint1})
 \cite{SamiDadich,InstantPreheating,FermionPreheating,VariableGravity,ala}.
The  effective mass of $\chi$ can be read of from  (\ref{Lint1}) as,
$m_{\chi} = g \left| \phi \right|$, \cite{InstantPreheating2}.
Production of
$\chi$ particles takes place after inflation provided $m_{\chi}$ changes non adiabatically,
\cite{QIRGWBSamiSahni,VariableGravity,ala,SamiDadich}
\begin{equation}
\dot{m}_{\chi} \gtrsim m_{\chi}^2 \longrightarrow \dot{\phi} \gtrsim g \phi^2.
\end{equation}
Let us now confirm that the above condition is met in the model under consideration. To this effect,
we estimate  $\dot{\phi}_{end}$  using the expression of slow roll parameter,
  $\epsilon = 4 \pi G \frac{\dot{\phi}^2}{H^2}$ ,
\begin{eqnarray}
\label{dotphiend}
 \dot{\phi}_{end}^2 \simeq V_{end} \to |\dot{\phi}_{end}| \cong V_{end}^{1/2}
\end{eqnarray}
 The condition for particle production then becomes,
\begin{align}
\left| \phi \right| &\lesssim \left| \phi_{prod} \right| \nonumber \\
&= \sqrt{\frac{\left| \dot{\phi}_{end} \right|}{g}} \nonumber \\
&= \sqrt{\frac{V_{end}^{\frac{1}{2}}}{ g}} \longrightarrow g^2 \gtrsim M_{pl}^{-4} V_{end}
\left( \phi_{pd} \leq M_{pl} \right), \label{phiprod}
\end{align}
Using Eq.~(\ref{phiprod}), we have following expression,
\begin{equation*}
\frac{\left| \phi \right|}{\left| \dot{\phi} \right|}
\approx \frac{\left| \phi_{prod} \right|}{\left| \dot{\phi}_{end} \right|}
= g^{- \frac{1}{2}} {\left| \dot{\phi}_{end} \right|}^{- \frac{1}{2}}.
\end{equation*}
The production time can be estimated\cite{VariableGravity, ala, SamiDadich, Starobinsky,Starobinsky2} as
\begin{align}
t_{pd} &\approx \frac{\phi}{\left| \dot{\phi} \right|}
\approx  g^{- \frac{1}{2}} {\left| \dot{\phi}_{end} \right|}^{- \frac{1}{2}}
\end{align}
  which allows us then to estimate wave number using uncertainty relation,
\begin{align}
k_{pd} \approx t_{pd}^{-1} \approx \sqrt{g \left| \dot{\phi}_{end} \right|}
\end{align}
The  occupation number for $\chi$ particles is given by the following expression,
\begin{align}
n_k \approx e^{- \frac{\pi k^2}{k_{pd}^{2}}}.
\end{align}
We can then find the number density and energy density of created particles,
\begin{align}
N_{\chi} &= \frac{1}{(2 \pi)^3} \int_{0}^{\infty}{n_K d^3 K} \nonumber \\
&\simeq \frac{{\left( g \left| \dot{\phi}_{end} \right| \right)}^{\frac{3}{2}}}{(2 \pi)^3}.
\end{align}
\begin{equation}
\rho_{\chi} = N_{\chi} m_{\chi}
\simeq \left[ \frac{{\left( g \left| \dot{\phi}_{end} \right| \right)}^{\frac{3}{2}}}{(2 \pi)^3} \right]
\left[ g \left| \phi_p \right| \right] .
\end{equation}
Now plugging $\left| \phi_p \right| \simeq g^{- \frac{1}{2}} {\left| \dot{\phi}_{end} \right|}^{\frac{1}{2}}$ in the above
equation, one finds,
\begin{equation}
\rho_{\chi} \simeq \frac{g^2 {\left| \dot{\phi}_{end} \right|}^2}{(2 \pi)^3} \label{rhochi}.
\end{equation}
Substituting $\left| \dot{\phi}_{end} \right| \approx V_{end}^{\frac{1}{2}}$ in the above equation, $\rho_{\chi}$ reduces to
\begin{equation}
\rho_{\chi} \simeq \frac{g^2 V_{end}}{(2 \pi)^3}.
\end{equation}
Using Eq.~(\ref{dotphiend}) and $\rho_{\phi}= \frac{\dot{\phi}^2}{2} + V(\phi)$,
\begin{equation}
(\rho_{\phi})_{end} \simeq \frac{3}{2} (\dot{\phi}^2 )_{end} \simeq \frac{3}{2} V_{end}\label{rhophiend}.
\end{equation}
Combining Eq.~(\ref{rhochi}) and Eq.~(\ref{rhophiend}) yields,
\begin{equation}
{\left( \frac{\rho_{\phi}}{\rho_{\chi}} \right)}_{end} \simeq \frac{3}{2} \frac{(2 \pi)^3}{g^2} \label{ratiorhophirhochi}.
\end{equation}
Assuming that the $\chi$ field gets converted into radiation and thermalization takes place instantaneously, we have,
\begin{equation}
\rho_r \approx \rho_{\chi}.
\end{equation}
We thus obtain,
\begin{equation}
{\left( \frac{\rho_{\phi}}{\rho_r} \right)}_{end} \simeq \frac{3}{2} \frac{(2 \pi)^3}{g^2}. \label{ratiorhophirhor}
\end{equation}
The nucleosynthesis constraint \cite{ala} dictates that
\begin{equation}
 \left(\frac{\rho_g}{\rho_r}\right)_{eq} \lesssim 10^{-2}, \label{rhogrhor1}
\end{equation}
where
\begin{equation}
 \left(\frac{\rho_g}{\rho_r}\right)_{eq}
 = \frac{64}{3 \pi} h_{GW}^2 \left(\frac{\rho_{\phi}}{\rho_r}\right)_{end} \label{rhogrhor2}
\end{equation}
Combining $h_{GW}^2 = \frac{H_{in}^2}{8 M_{pl}^2}$ and $H_{in}^2 = \frac{V_{in}}{3 M_{pl}^2}$ with Eq.~(\ref{rhophiend}),
Eq.~(\ref{rhogrhor1}) and Eq.~(\ref{rhogrhor2}), one obtains the following inequality,
\begin{equation}
 (\rho_r)_{end} \gtrsim  \frac{400 }{3 \pi} \frac{V_{in} V_{end}}{M_{pl}^4}. \label{ineqrhor}
\end{equation}
Considering the form of our potential at the beginning and end of inflation, the above inequality reduces to,
\begin{equation}
 (\rho_r)_{end} \gtrsim  \frac{400 }{3 \pi} \frac{V_{0}^2 }{M_{pl}^4 \cosh{(\beta^n \chi_{in}^n)} \cosh{(\beta^n \chi_{end}^n)}}.
\end{equation}
Using $M_{pl} = 2.4 \times 10^{18} GeV$ together with $\chi_{end} = 0.88$, $\chi_{in} = 0.44$ and
$V_0= 4.64 \times 10^{60} GeV^4$, where these values are obtained for $n=6$ and $\beta=1$, the following bound
on $(\rho_r)_{end}$ is calculated,
\begin{equation}
 (\rho_r)_{end} \gtrsim 2.48 \times 10^{49} GeV^4.
\end{equation}
Using $(T_r)_{end} \sim [(\rho_r)_{end}]^{1/4}$, the bound on $(T_r)_{end}$ can be written as
\begin{equation}
 (T_r)_{end} \gtrsim 2.2 \times 10^{12} GeV. \label{ineqTrend}
\end{equation}
Using Eq.~(\ref{rhophiend}), the inequality  Eq.~(\ref{ineqrhor}) can be rewritten as
\begin{equation}
 \left(\frac{\rho_{\phi}}{\rho_r}\right)_{end} \lesssim \frac{9 \pi M_{pl}^4}{800 V_{in}} =
 \frac{9 \pi M_{pl}^4 \cosh{(\beta^n \chi_{in}^n)}}{800 V_{0}}.
\end{equation}
Evaluating the  expression on the right hand side for $\beta=1$ and $n=6$ as above, the above inequality is calculated as,
\begin{equation}
 \left(\frac{\rho_{\phi}}{\rho_r}\right)_{end} \lesssim 2.53 \times 10^{11}. \label{ineqnumber}
\end{equation}
Plugging Eq.~(\ref{ratiorhophirhor}) in Eq.~(\ref{ineqnumber}), we obtain the bound on $g$,
\begin{equation}
 g\gtrsim 3.8 \times 10^{-5}. \label{bound_g}
\end{equation}
 We now calculate the bound on $h$. For the reaction,
 $\chi \rightarrow \bar{\psi} \psi$, the decay width $\Gamma_{\bar{\psi} \psi}$ satisfies the following inequality
 \cite{LDH,BPS,Starobinsky,Starobinsky2,ala,VariableGravity,SamiDadich,FermionPreheating,Campos1,Campos2,
 InstantPreheating,InstantPreheating2,QIRGWBSamiSahni},
 \begin{eqnarray}
  \Gamma_{\bar{\psi} \psi} \gg H_{end} \rightarrow h^2 \gtrsim \frac{8 \pi H_{end}}{g |\phi|}  .
 \end{eqnarray}
Now using $H^2 = \frac{8 \pi G}{3} \rho_{\phi}$, Eq.~(\ref{rhophiend}) and  $|\phi| \leq M_{pl}$ for our potential one obtains
  \begin{eqnarray}
   h^2 \gtrsim  \frac{ 4 \pi \sqrt{2 V_{end}}}{g M_{pl}^2} = \frac{ 4 \pi \sqrt{2 V_0/(\cosh{(\beta^n \chi_{end}^n)}) }}{g M_{pl}^2} ,
  \end{eqnarray}
which simplifies to
\begin{eqnarray}
 h \gtrsim \frac{2.5 \times 10^{-3}}{\sqrt{g}} \label{bound_h},
\end{eqnarray}
for $n=6$ and $\beta=1$.  From Eq.~(\ref{bound_g}),  $g$ can take any value greater than or equal to $3.8 \times 10^{-5}$ and based on the value taken by
 $g$, $h$ satisfies  Eq.~(\ref{bound_h}). The bounds on $g$ and $h$ that are calculated above are such that $(T_{r})_{end}$ always
 satisfies the inequality (\ref{ineqTrend}). 
 
 \vspace{.5cm}
 \chapter{\hspace{.2cm}\bf Appendix B: Relic Gravity Wave Spectrum} \label{appxB}
\vspace{.5cm}
One of the tests for inflationary models is the measurement of the spectrum of gravity waves. In this appendix, considering the given
reheating temperature, we estimate the spectrum of the relic gravity wave.  The gravitational wave equation in flat FRW space time can be written  in its standard form 
as,
\begin{equation}
h^{''}_k (\tau) + 2 \frac{a'}{a} h^{'}_{k} (\tau) + k^2 h_k (\tau) = 0.
\end{equation}
This equation describes how gravity wave  evolves with time in a flat FRW Universe. The energy spectrum is defined as \cite{Boyle:2005se},
\begin{equation}
\Omega_{gw} (k, \tau) = \frac{1}{\rho_{crit} (\tau)} 
\frac{d \left( \left< 0 \left| \hat{\rho}_{gw} (\tau) \right| 0 \right> \right)}{ d \left( ln k \right)},
\end{equation}
where $\rho_{crit} = \frac{3 H^2 (\tau)}{8 \pi G}$ and the gravitational energy density $\rho_{gw}$ is given by
\begin{equation}
\rho_{gw} = - T^{0}_{0} = \frac{1}{64 \pi G} 
\frac{{\left( h'_{ij} \right)}^2 + {\left( \nabla h_{ij} \right)}^2}{a^2}.
\end{equation}
It can be shown that the spectrum of gravity waves depends on the  equation of state of the dominant fluid comprising the universe \cite{RGWBISamiSahni,Boyle:2005se}. 
Therefore, the gravitational waves can be categorized according to  three epochs of the universe -  the matter dominated regime,
the radiation dominated regime and the kinetic regime, and in these regimes the energy spectrum of relic gravity waves is given by the following expressions \cite{RGWBISamiSahni}:

\begin{align}
&\Omega_g^{(M.D.)} (\lambda) = \frac{3}{8 \pi^3} h_{G.W.}^2 \Omega_{om} { \left( \frac{\lambda}{\lambda_h} \right) }^2, \label{wavematt}
&\lambda_{M.D.} < \lambda \leq \lambda_{h} \\ 
&\Omega_g^{(R.D.)} (\lambda) = \frac{1}{6 \pi} h_{G.W.}^2 \Omega_{or}, &\lambda_{R.D.} < \lambda \leq \lambda_{M.D.} \label{waverd}\\ 
&\Omega_g^{(kin)} (\lambda) = \Omega_g^{(R.D.)} \left( \frac{\lambda_{R.D.}}{\lambda} \right),
&\lambda_{kin} < \lambda \leq \lambda_{R.D.} , \label{wavekin}
\end{align}
where, $\lambda_{h} = 2 c H_{0}^{-1} \approx 1.8 \times 10^{28} h^{-1} \mbox{cm} \approx 2.57 \times 10^{28} \mbox{cm} $. 
$\lambda_{M.D.}$, $\lambda_{R.D.}$  can be estimated using the boundary conditions: $\Omega_g^{(M.D.)} |_{\lambda = \lambda_{M.D.}} = \Omega_g^{(R.D.)} |_{\lambda = \lambda_{M.D.}}$
and
$\Omega_g^{(kin)} |_{\lambda = \lambda_{R.D.}} = \Omega_g^{(R.D.)} |_{\lambda = \lambda_{R.D.}}$. So, we have,
\begin{align}
\lambda_{M.D.} &=\frac{2 \pi}{3} \lambda_h {\left( \frac{\Omega_{or}}{\Omega_{om}} \right)}^{\frac{1}{2}} \\
\lambda_{R.D.} &=4 \lambda_h {\left( \frac{\Omega_{or}}{\Omega_{om}} \right)}^{\frac{1}{2}} 
\frac{T_{M.D.}}{T_{rh}},
\end{align}
and $\lambda_{kin} $ is given by
\begin{align}
\lambda_{kin} = c H^{-1}_{kin} \left( \frac{T_{rh}}{T_0} \right) { \left( \frac{H_{kin}}{H_{rh}} \right) }^{\frac{1}{3}}.
\end{align}
The gravity waves with wavelength $\lambda < \lambda_{R.D.}$ are generated during the  kinetic regime
($\omega \sim 1 $). The spectrum of these waves is inversely proportional to the wavelength. During the radiation phase,
$\lambda_{R.D.}< \lambda<\lambda_{M.D.}$, it is constant. However, for $\lambda > \lambda_{M.D.}$, the waves are created in the 
matter phase and its spectrum increases according to Eq. (\ref{wavematt}). Now we calculate $\lambda_{M.D.}$, $\lambda_{R.D.}$ and
$\lambda_{kin}$.
\begin{align}
\lambda_{M.D.} &=\frac{2 \pi}{3} \lambda_h {\left( \frac{\Omega_{or}}{\Omega_{om}} \right)}^{\frac{1}{2}} 
= 3.10607 \times 10^{26} \mbox{cm}   .
\end{align}
Here, we plugged $10^{-5}$ and $0.3$ for $\Omega_{or}$ and $\Omega_{om}$ respectively.
\begin{align}
\lambda_{R.D.} &=4 \lambda_h {\left( \frac{\Omega_{or}}{\Omega_{om}} \right)}^{\frac{1}{2}} 
\frac{T_{M.D.}}{T_{rh}}= 3.64851 \times 10^{5} \mbox{cm},   
\end{align}
where, we used $T_{M.D.}=1.3524 \mbox{eV}$ and $T_{rh}=2.2 \times 10^{12} \mbox{GeV}$.
\begin{equation}
\lambda_{kin} = c H^{-1}_{kin} \left( \frac{T_{rh}}{T_0} \right) { \left( \frac{H_{kin}}{H_{rh}} \right) }^{\frac{1}{3}}.
\end{equation}
Using $H_{kin} \approx H_{rh} \approx H_{end}$ and $H_{end}^{-1} = \left( 4 \pi G V_{end} \right)^{\frac{1}{2}}$ we get,
\begin{equation}
\lambda_{kin} = \frac{c}{\sqrt{4 \pi G V_{end}}} \times \left( \frac{T_{rh}}{T_0} \right). \label{lambdakin}
\end{equation}
We have $n = 6$, $\beta = 1$, $\chi_{end} = 0.88$, $V_0 = 4.64 * 10^{60} GeV^4$,  
\begin{align}
V_{end} &= \frac{V_0}{\cosh{\left( \beta^n \chi_{end}^n \right)}} = 4.18 \times 10^{60} GeV^4 . 
\end{align}
  
Substituting $V_{end}$, $T_0 =2.34813 \times 10^{-13} \mbox{GeV}$ and $T_{rh}=2.2 \times 10^{12} \mbox{GeV}$
in Eq. (\ref{lambdakin}), $\lambda_{kin}$ turns out to be,
\begin{equation}
\lambda_{kin} =  0.306992 \ \mbox{cm}.
\end{equation}
Now using $h_{G.W.}^2 = \frac{H_{in}^2}{8 M_{pl}^2}$ and $H_{in}^2 = \frac{V_{in}}{3 M_{pl}^2}$ one obtains,
\begin{equation}
h_{G.W.}^2 = \frac{V_0}{24 M_{pl}^4 \cosh \left( \beta^n \chi_{in}^n \right)}.
\end{equation} 
Using $\beta = 1$, $\chi_{in} = 0.44$, $V_0 = 4.64 \times 10^{60} GeV^4$ and $M_{pl} = 2.4 \times 10^{18} GeV$,
$h_{GW}^2$ can be written as
\begin{align}
h_{G.W.}^2 = 5.82139 \times 10^{-15},
\end{align}
which simplifies Eqs. (\ref{wavematt}), (\ref{waverd}) and (\ref{wavekin}) as following,
\begin{align}
\Omega_g^{(M.D.)} (\lambda) &= 3.19789 \times 10^{-74} \lambda^2, \ \nonumber \\ 
& \mbox{for} \ \ 3.10 \times 10^{26} \mbox{cm} < \lambda \leq 2.57 \times 10^{28} \mbox{cm} , \\
\Omega_g^{(R.D.)} (\lambda) &= 3.08834 \times 10^{-21}, \ \nonumber \\ 
& \mbox{for} \ \ 3.64851 \times 10^5 \mbox{cm} < \lambda \leq 3.10 \times 10^{26} \mbox{cm} , \\
\Omega_g^{(kin)} (\lambda) &= \frac{1.12678 \times 10^{-15}}{\lambda}, \ \nonumber \\
& \mbox{for} \ \ 0.306992 \mbox{cm} < \lambda \leq 3.64851 \times 10^5 \mbox{cm}.
\end{align}

\end{appendices}



\end{document}